\documentstyle[12pt]{article}
\begin{document}
\title{Cell Division, Differentiation and Dynamic Clustering}
\author{
        Kunihiko Kaneko\\
        {\small \sl Department of Pure and Applied Sciences}\\
        {\small \sl University of Tokyo, Komaba, Meguro-ku, Tokyo 153, JAPAN}\\
       and\\
        Tetsuya Yomo\\
        {\small \sl Department of Biotechnology}\\
        {\small \sl Faculty of Engineering }\\
        {\small \sl Osaka University 2-1 Suita, Osaka 565, JAPAN}
\\}
\date{}
\maketitle
\begin{abstract}
A novel mechanism for cell differentiation is proposed,
based on the dynamic clustering in a globally coupled chaotic system.
A simple model with metabolic reaction, active transport of chemicals
from media, and cell division is found to show three successive stages
with the growth of the number of cells; coherent growth,
dynamic clustering, and fixed cell differentiation.  At the last stage,
disparity in activities, germ line segregation, somatic cell differentiation,
and homeochaotic stability against external perturbation
are found.  Our results, in consistent with the experiments
of the preceding paper, imply that cell differentiation
can occur without a spatial pattern.  From dynamical systems viewpoint,
the new concept of ``open chaos" is proposed, as a novel
and general scenario for systems with growing numbers of elements,
also seen in economics and sociology.
\end{abstract}
\section{Introduction}
Why do cells differentiate?  The orthodox answer to this question
is that the mechanism is completely determined by genetic codes.
This belief is widely accepted by most molecular biologists.
Is it correct though ? It is known that genes are not changed in the
course of cell differentiation\cite{Gurdon}.
Cells with identical genes can differentiate, even when in
the same environment.
Hence it is not a trivial question how identical
sets of genes can produce a variety of different cells,
not only from a biological but also from a dynamical systems viewpoint.
Experiments on cell differentiation of E.~Coli by one of
the authors (TY),  however, leads to a serious question to this
widely accepted answer.
Indeed, as reported in the preceding paper \cite{Yomo},
cells with identical genes may split into several groups
with different enzymatic activities.
Even prokaryote cells with
identical genes can be differentiated there.
We note that these
cells are under liquid culture, thus are in an identical environment.
Of course, spatial information is also important in differentiation.
However, usually one does not seriously discuss whether a spatial pattern is
necessary for the cell differentiation.  Often,
differentiation and pattern formation are discussed
together without distinction.  The experimental results of the
preceding paper, however, suggest that differentiation can occur
without such spatial (positional) information.
In the present paper we demonstrate theoretically that cell differentiation
is possible without spatial information.
Here we note that cells
interact with the environment, which is affected by all the other cells.
All cells interact with other cells through the environment.
The purpose of the present paper is to present an alternative answer
to the top question, with an emphasis on the cell-environment interaction.
Our answer is based on dynamic clustering in a globally coupled dynamical
system.
One of the simplest examples of such clustering is given by
globally coupled dynamical systems \cite{KK-GCM}.
When many identical elements with chaotic dynamics
interact globally through a mean field, it is found that
the elements differentiate into some clusters. In each cluster,
elements oscillate synchronously, while elements in a different cluster
oscillate with a different phase, frequency, or amplitude.
Thus spontaneous differentiation of elements is possible through interaction
among chaotic elements.
In the present paper we extend this idea of clustering to
spontaneous cell differentiation by introducing
a simple metabolic chemical reaction and cell division dynamics.
The clustering mechanism, although it provides a universal key to
differentiation, is not enough for the explanation of fixed differentiation
in multi-cellular organisms.  Indeed, differentiation by clustering
is temporal in nature:  A group of cells with rich substrates at one time
turns to be poor after some time in accordance with the oscillation.
By introducing a model with a growing number of cells,
we find that the appearance of
a new stage after this dynamic clustering, where the
differentiation of cells is fixed.
Disparity in activities emerges.
Thus our result will provide a novel clustering behavior, also
from the viewpoint of dynamical systems.
As a dynamical system, our problem has one
important and innovating feature: growth of the dimension of
phase space with cell divisions.  Since the chemical dynamics in
each cell is represented by a set of differential equations of a given
number of degrees of freedom, the total number of
degrees of freedom of the system
increases with the number of cells $N(t)$.  In our system,
the increase is closely associated with chaotic instability.
Our model of cell growth consists of
(i) nonlinear dynamics in each cell (ii) nonlinear and global
interaction among cells through a medium
(iii) growth and death of a cell depending
on its internal state.  Such processes can often be seen in a system
with growth, by reinterpreting a cell as any replicating unit.
For example, economic developmental processes have
the above three features, by regarding resources and money as
the chemicals and activities.  Emergence of spontaneous differentiation
in our cell society, thus, can be related with the
sharing of resources, division of labor, and the
formation of classes in an economical or sociological system.
The present paper is organized as follows.
A specific model is introduced in \S 2, combining
the processes of metabolic reaction, active transport of chemicals
into cells, cell division, and death.  Numerical results of
the model, given in \S 3, show three stages of differentiation;
coherent growth, dynamic clustering, and fixation of differentiation.
Detailed dynamical aspects are presented.  In \S 4, implications of
our results to cell differentiation are given.  The origin of germ line
segregation,
differentiation of somatic cells in multicellular organisms,
the possible mechanism of programmed death,
and homeochaotic stability against external perturbation are discussed.
In \S 5, we point out a new mechanism of chaotic instability
in a system with a growing number of degrees of freedom.
\section{Model}
The biochemical mechanisms of cell growth and division are very complicated,
and include a variety of catalytic reactions.  The reaction
occurs both at the levels of inter- and intra- cells.
Hence it is almost impossible to construct
a complete descriptive model.  Here we introduce a simpler model which
captures the essence of these processes (see Fig.1 for a schematic illustration
of our model).
\vspace{.1in}
---------Fig.1   ---------
\vspace{.1in}
In the model we have to include the following processes:
\begin{itemize}
\item { Metabolic Reaction within each Cell : Intra-cell Dynamics}
\item { Interaction with Other Cells through Media: Inter-cell  Dynamics}
\item { Cell Division}
\item {Cell Death}
\end{itemize}
{\bf (A) Chemicals}
 As a set of dynamical variables we need some chemicals' concentrations
in each cell, and also those of the medium surrounding the cells.  Here we
use the following variables; a set of concentrations of chemical substrates
$x^{(m)} _i(t)$, the concentration
of $m$-th chemical species at the $i$-th cell, at time $t$.  The
corresponding concentration of the species in the medium is denoted as
$X^{(m)} (t)$.  We assume that the medium is
well stirred, and neglect the spatial variation of the concentration.
Furthermore we regard the chemical species
$x^{(0)}$ ( or $X^{(0)}$ in the media) as
playing the role of the source for other substrates.  Another assumption we
make
here is the existence of enzymes (for convenience  and simplicity
we assume that there is a corresponding enzyme $E^{(m)}$ for each chemicals
$x^{(m)}$).
{\bf (B) Metabolic Reaction}
The metabolic reaction here is schematically shown in Fig.1 b).
In the present paper we make further simplifications:
(1) Only three chemicals including the source $x^{(0)}$ are considered.
In other words, each cell contains
the variables $x^{(0)}_i (t)$, $x^{(1)}_i (t)$, and $x^{(2)}_i (t)$.
Of course, this is a vast simplification.  There can be
several orders of cascade reactions in the course of synthesis of
DNA \cite{TheCell}.  We will see, however, that our simple reaction scheme is
enough for our new scenario of cell differentiation.
(2) Simplification of enzyme dynamics:
To be specific, we take the reaction scheme shown in Fig.1b).
Here the enzyme $E^0$ is constitutive while the others, $E^1$ and $E^2$, are
inductive  \cite{TheCell2}.
First we assume that the concentration of the enzyme $E^{(0)}$ is constant,
given by $E^0$, for simplicity.  Here, the generation of the enzymes
$E^{(1)}$ and $E^{(2)}$ are activated by the chemical $x^{(2)}$.
The rate equation within each cell (including the enzymatic dynamics) is
written as:
\begin{equation}
dx^{(1)}_i (t)/dt  = E^{(2)}_i (t)x^{(2)}_i (t)- E^{(1)}_i (t)x^{(1)} (t)+
E^0 x^{(0)}_i (t).
\end{equation}
\begin{equation}
dx^{(2)}_i (t)/dt  = E^{(1)}_i (t)x^{(1)}_i (t)- E^{(2)}_i (t)x^{(2)}(t)-
 \delta \times x^{(2)}_i (t).
\end{equation}
The dynamics of enzymes  $E^{(m)}_i (t)$ can be highly complex and nonlinear,
by gene expression, transcription, translation, and modification
\cite{TheCell}.  Following the scheme in Fig. 1b), we adopt here
\begin{math}
dE^{(1)}_i (t)/dt  = a_1(x^{(2)}_i (t) -d_{e1} E^{(1)}_i (t));
\end{math}
\begin{math}
dE^{(2)}_i (t)/dt  = a_2(x^{(2)}_i (t) -d_{e2} E^{(2)}_i (t)).
\end{math}
as one of the simplest choices.
Each enzyme is created from chemicals by reactions within
a cell. We neglect the possibility of transportation of enzymes across
cells ( since the sizes of enzymes are much larger).
For simplicity,
we adiabatically solve
the equations to get $E^{(1)}_i (t) =e_1 x^{(2)}_i (t)$,
and $E^{(2)}_i (t) =e_2 x^{(2)}_i (t)$, with
constants $e_1$ and $e_2$.
Throughout the paper we adopt this elimination.  Further simplification we make
here is the neglection of the reaction $x^{(2)} \rightarrow x^{(1)}$.
Thus the terms with $E^{(2)}$ are neglected, although some simulations
with these terms lead to qualitatively identical results.
{\bf (C) Active Transport and Diffusion through Membrane}
A cell takes chemicals from the surrounding medium.
The rates of chemicals transported into a cell are proportional to their
concentrations outside.
Further we assume that this transport rate also depends nonlinearly on
the internal state of a cell.  Since the transport here requires
energy \cite{active}, the transport rate depends on the
activities of a cell.  The rate can depend nonlinearly on the
chemicals in the cell.  To be specific, we choose the following
form;
\begin{equation}
Transp^{(m)}_i (t) = p(\sum_{k=1}^2 x^{(k)}_i (t))^{\nu } X^{(m)} (t)
\end{equation}
where $\nu$ is taken to be 3 throughout the paper, although
other  nonlinear dependences ( $\nu >1$, i.e., with positive feedback effect)
lead to qualitatively similar results.  The summation
$(\sum_{k=1}^2 x^{(k)}_i (t))$ is introduced here
to mean that a cell with more chemicals is more active.
This form, again, is rather arbitrarily chosen, but
similar forms with "active" transport can lead to the same result.
Besides the above active
transport, the chemicals spread out through the membrane with normal
diffusion by
\begin{equation}
Diff^{(m)}_i (t) = D(X^{(m)} (t) - x^{(m)}_i )
\end{equation}
Combining the processes (B) and (C),
the dynamics for $x^{(m)}_i (t)$ is given by
\begin{equation}
dx^{(0)}_i (t)/dt  = -E^0 x^{(0)}_i (t)+ Transp^{(0)}_i (t)+Diff^{(0)}_i (t) ,
\end{equation}
\begin{equation}
dx^{(1)}_i (t)/dt  = E^0 x^{(0)}_i (t)- e_1 x^{(2)}_i (t)x^{(1)}_i (t)+
Transp^{(1)}_i (t)+Diff^{(1)}_i (t) ,
\end{equation}
\begin{equation}
dx^{(2)}_i (t)/dt  = e_1 x^{(2)}_i (t)x^{(1)}_i (t) -\delta \times x^{(2)}_i
(t)+
Transp^{(2)}_i (t)+Diff^{(2)}_i (t) ,
\end{equation}
Since the processes (B) here are just the transportation of
chemicals through membranes, the
sum of the chemicals must be conserved.  The dynamics of the chemicals in the
medium is then obtained by converting the sign, i.e.,
\begin{equation}
dX^{(m)}(t)/dt  =-\sum_{i=1}^N \{ Transp^{(m)}_i (t) +  Diff^{(m)}_i (t) \}.
\end{equation}
Since the chemicals in the medium
can be consumed with the flow to the cells, we need some flow of chemicals
(nutrition) into the medium from the outside of the container.  Here
we assume that only the source chemical $X^0$ is supplied by a flow
into the container. By denoting the external concentration of the chemicals
by $\overline{X^0}$ and its flow rate per volume of
the tank by $f$, the dynamics of source chemicals in the
media is written as
\begin{equation}
dX^{(0)}(t)/dt  =f(\overline{X^0} -X^0)-\sum_{i=1}^N \{ Transp^{(0)}_i (t) +
Diff^{(0)}_i (t) \}.
\end{equation}
{\bf (D) Cell Division}
Through chemical processes, cells can replicate.  For the
division, accumulation of some chemicals is required.
In our model the final product from the chemical species ``2" is assumed to act
as the chemical for the cell division ( note the term $-\delta \times x^{(2)}
_i (t)$
in eq. 2).  For example assume that chemical $2$ is a mono-nucleotide.
Then the DNA synthesis process occurs through the reaction $2 \rightarrow DNA$
in Fig.1; thus $\frac{d DNA}{dt} \propto x^{(2)}$. Accordingly we impose the
following condition for cell division: the $i$-th cell
(born at time $t=t_0(i)$) divides into two
at a time $T$ such that
\begin{equation}
\int _{t_0 (i)}^T dt x^{(2)} _i (t) > R
\end{equation}
is satisfied, where $R$ is the threshold for cell replication.
Here again, choices of other similar division
conditions can give qualitatively similar results as those to be discussed.
The essential part is that the division condition satisfies
an integral form representing the accumulation of DNA as in eq. (10).
When a cell divides, two almost identical cells are formed.
The chemicals $x^{(m)}_i $ are almost equally distributed.
"Almost" here means that each cell after a division has
$\frac{x^{(m)}_i}{2}+\epsilon $ and $\frac{x^{(m)}_i}{2}-\epsilon $
respectively with a small "noise" $\epsilon$, a random number with
small amplitude, say over
 $[-10^{-3}, 10^{-3}]$.  We should note that this inclusion of imbalance is
not essential to our differentiation.  Indeed any tiny difference
is amplified to yield a macroscopic differentiation.
{\bf (E) Cell Death}
To avoid infinite growth,  a condition for cell death is further imposed.
Here we choose either a deterministic or a probabilistic death mechanism.
In the former case, we choose the condition for the death
as
\begin{equation}
\sum_{j=1}^{2} x^{(j)}_i (t) < S,
\end{equation}
where $S$ is a threshold for ``starvation".
This choice is again rather arbitrary.  We have assumed that
a cell dies when the chemicals included therein are too little.
Again, a choice of similar other forms is expected to give the same results.
Here, the chemicals inside the dead cells are released into
the medium.  Thus there is a jump in $X^{(j)} (t)$ at every cell death.
In the probabilistic case, a number of randomly chosen cells
(together with the chemicals included therein) are removed
per given time steps (decimation).  This choice is closer
to experimental situations, since incuvated cells
are decimated per some time steps in order to avoid the divergence
of the number of cells (see the preceding paper).
In fact, both the deterministic and the probabilistic deaths
give qualitatively the same behavior with regard to the three stages to
be discussed.
\section{Results: Three Stage Differentiation}
A typical example for the growth of the number of cells
is plotted in Fig.2,  as well as the overlaid oscillation
of chemicals $x^{(2)}_i (t)$ over all cells in Fig.3 .
In Fig.2, the number of cells doubles
at certain time during the first stage, while the number increases almost
linearly
with time at the last stage.  As can be seen in Fig.3,
we have observed the following three stages of evolution,
with the growth of the number of cells, for
a wide range of parameters.
\vspace{.1in}
------Fig.2 ----
------Fig.3 ----
\vspace{.1in}
(a) Stage I: Coherent growth
All the metabolic reactions of cells oscillate coherently.
Thus cells grow coherently.
Starting from a single cell,  the number increases as
$1 \rightarrow 2 \rightarrow 4 \rightarrow 8 \rightarrow 16 \cdots$
at the stage I.
(b) Stage II: Dynamic clustering
If there is intra- and/or inter-cell nonlinear dynamics, the
coherent oscillation can lose its stability as the cell number increases.
As can be seen in Fig. 3 of the overlaid
time series of $x^{(2)} _i (t)$,
the variance of the cells' oscillations is enhanced with time.
Then dynamic clustering sets in.
Depending on the parameters,  the number of clusters can be different.
As the diffusion constant $D$ is decreased, the number of clusters
increases, and the oscillation gets more complicated.  The projected
orbit of $(x^{(1)}_i (t) ,x^{(2)}_i (t))$ is given in Fig.4 a).
The phases of oscillations, as well as their amplitudes
vary by cells.  Some cells start to have large concentrations of chemicals,
which prepares them for the differentiation at the next stage.
The origin of the clustering (i.e., temporal differentiation) lies in
the instability of the reaction and transport dynamics.  Any tiny
difference between two cells is amplified to a macroscopic difference.
The mechanism of clustering has been investigated in globally coupled
maps \cite{KK-GCM}, although here
there exists a novel feature in dynamical systems,
as will be discussed in \S 5.
We note that the clustering here is possible by the interaction among cells
through the chemicals in  the medium, whose concentrations can also oscillate
in time.
\vspace{.1in}
----------fig.4   --------------
----------fig.5   --------------
\vspace{.1in}
(c) Stage III: Fixation of cell differentiation:
As the number of cells increases further, our cell society
enters into a new stage.  An example of overlaid time series is
given in Fig.3, as well as the orbit in phase space
$(x^{(1)} _i (t),x^{(2)} _i (t))$ in Fig.4b). We note that the cells are
classified into two
completely distinct groups; the population of cells in one group is
very few (one or two in the example) but they
contain large concentrations of chemicals, while
the population of the other group is large but the concentrations of the
chemicals are much smaller.  We call the former cells
{\sl active} and the latter as {\sl sleeping}. The active cells
replicate much faster than the other (sleeping) ones.
In Fig. 4b), orbits with a large amplitude are those of the
active group. The segment of straight lines at the upper middle
is a result of the cell division.
(Here $x^{(2)}$ of the sleeping cells is so small that their orbits in
the $(x^{(1)} _i (t),x^{(2)} _i (t))$ plane are hardly visible in Fig. 4b)).
In the extreme, and here, typical case the number of cells in the active group
is just one.  It divides almost periodically in time.  After each division,
one of the two created cells
takes more chemicals than the other.  The difference between the two cells
is enhanced, till the weaker one belongs to the sleeping group.
Thus the number of cells increases by one per some period as long
as the "cell death" condition is not satisfied.
The sleeping cells, on the other hand, are not identical. They are again
weakly differentiated
by the concentrations of chemicals ( see Fig.5 ).  This differentiation
depends on the ordering of the time of birth of the cell.
Temporally there are weak oscillations in the chemicals of a sleeping cell.
The above regular growth is seen if $\delta $ ( production rate
of DNA from $x^{(2)}$) and $D$ ( diffusion coupling with a medium)
are large ( say $\delta >.2$, $D>.08$).
When the parameters $\delta$ and/or $D$ are smaller,
the dynamics is more complicated: the number of active cells is larger than
one (stage ``complex-III").  Their
oscillations are not periodic, and the division cycle is not regular either.
See Fig.6a) for the irregular growth of cells, while the overlaid time
series of $x^{(2)}_i (t)$ are given in Fig. 6b), as well as the plot of
the $(x^{(1)}_i , x^{(2)}_i)$ in Fig. 6c).  Here, the oscillations of
active cells are chaotic and display dynamic clustering.
By a division of a cell in either group, the balance of the number of cells
between the two groups may be destroyed.  After this (rare) occurrence,
one (or few) of the cells in one group switch to the other group.
In Fig.7, sleeping cells "wake up" around $t=4620$ and get active by taking
chemicals from the medium, while active ones get inactive.
Such waking-up of sleeping cells has also been observed
in the experiments of the preceding paper.
Before closing the paragraph on stage III, we note that for some
parameter regimes, ( e.g., much larger $D$),
this stage has not yet been observed, and
the cell society remains at stage II.
\vspace{.1in}
----------Fig.6  a)b)c)---------------
----------Fig.7   --------------------
\vspace{.1in}
(d) Cell Death
After the number of cells gets large enough, the external
source term is not sufficient to support any more cells.
The number saturates at the maximal number
and fluctuates around it.
When there is just one active cell, the cell still divides
periodically in time.  However, one of the divided cells dies
within the same period.  Thus the number of cells does neither increase nor
decrease.
Cell linkage for this case is shown in Fig. 8.  The cell linkage diagram
shows from which cell a new cell is born at time $t$ (given by
the vertical axis).
In Fig.8, we can clearly distinguish active and sleeping cells (
where the cell index $\ell$ (horizontal axis) satisfies $\ell \leq 28$ or
$\ell \geq 103$).  Successive division and death is seen for the cells
with the index $28<\ell<103$.
\vspace{.1in}
-------------Fig.8  -----------------
\vspace{.1in}
If we wait for a very long time, the division condition is
satisfied for sleeping cells also (since the condition is of an intergal form).
Then divisions of many cells occur successively within a short time scale, but
many of the divided cells die right after the divisions.  An almost
simultaneous death of many cells occurs.  Thus the number returns to
the level before the multiple division.
If the number of active cells is larger than one
(``complex-III stage"), the death process is
irregular, while the total number of cells is constant on average.
Here switching between active and sleeping cells can occur through
cell death.  When the number of active cells is reduced,
the cell society goes back to the stage II, where many cells
with similar activities compete for resources, and show
dynamic clustering.  After some duration of stage II,
few cells become
active, and the society comes back to the complex-III stage.
Accordingly temporal oscillation is observed in the ratio
between active and sleeping cells.
\section{Implications to Cell Differentiation}
The results in the previous section have many implications for
cell growth and differentiation.
\begin{itemize}
\item{ Explanation of the results in the preceding paper}
Since the present toy cell with simple metabolic reaction systems
can show dynamic clustering, it is rather natural to
assume that the present clustering can appear in an experimental
system.  In the experiment of the preceding paper, differentiation
of cells and dynamic clustering of oscillations of E-coli are
observed.  The possibility of fixation of differentiation corresponding to
our stage III is also suggested experimentally.
Furthermore, the temporal oscillation of the populations
of active and sleeping cells is found experimentally, corresponding
to the complex-III stage in our simulation.
We note that the differentiation
in the experiment is performed with the use of liquid culture,
where the coupling is global (through a medium) as in our model.
Thus cell differentiation occurs even without a spatially local
interaction, in strong contrast with conventional models
for differentiation and pattern formation ( e.g., reaction-diffusion
equation model of Turing-type or a cellular automaton model).
One might think that the deterministic death condition is different
from the previous experimental situation.  In the experiments, some cells are
randomly removed to avoid overgrowth in the medium.  To take the
experiments into account, we have also simulated a model with
stochastic decimation.  The results are essentially the same
as those given so far, obtained by the deterministic death condition.
Of course, death is
not periodic, thus the growth is not completely periodic in the
regular regime of the deterministic case.  Still the dynamical behavior
is the same except for some effects of noise.
Statistically the ratio of active to
sleeping cells is kept at the same level.  When an active cell is eliminated,
for example, one of the sleeping cells starts getting more chemicals and
becomes an active cell.
\item{ From time sharing to emergence of disparity in activities}
In the present model, many cells compete with each other
for finite resources ( the source term $X^{(0)}$).
Generally speaking, coherent growth may not be advantageous in
a system with finite resources, since all elements need
the same amount of resources at the same time. There can be
two remedies to this problem; the time sharing system and
disparity of activities by elements.
In the former case, time sharing ( also often adopted in computer systems),
is accomplished here through temporal differentiation (clustering).
Through clustering, elements use resources successively,
thus avoiding strong competition for the resources.
In the latter case, elements split into (two) fixed groups.
One group is ``rich" and uses the resources more
and grows much faster than the other (poor) group.
In our system, these two types of differentiation appear
as successive stages.
We should note that the active and sleeping cells coexist.
Sleeping cells, although they may look like a ``loser" for
the competition for resources, can live together with
active ones.  Indeed this coexistence is
important for the stability of cell history and society
to be discussed.
When there are many active cells, they compete with each other
for resources.  In this case, the oscillations are
chaotic and show dynamic clustering.  Thus time sharing
of resources is still adopted among active cells.
\item{ Germ line segregation}
To have a stable cell society, cells, once differentiated, often must be fixed
in time. As found at stage III, such fixation occurs in our simulation.
This splitting into active and sleep cells
reminds us of the germ line segregation.
Germ cells are distinguished from somatic cells by their very high
division ability.
The emergence of the  germ line segregation in
our simple model is rather striking.
\item{ Somatic Cell Differentiation}
At stage III,
concentrations of chemicals in the sleeping group ( somatic cells)
are again differentiated into smaller groups.
This differentiation is according to the slight difference
of the amount of chemicals, in our model.  We should note that a very
simple metabolic network with three chemicals has lead to this
differentiation.  It is expected that the inclusion of more
chemicals can easily provide
a larger variety of differentiated cells as observed in a real cell society.
In our model, the growth speed of
somatic cells is very low.  This observation agrees rather well
with the well-known fact in biology; that
the division speed of cells gets much slower,
when they are differentiated \cite{diff-slow}.
\item{ Sleep/Active Switch and Homeochaos}
Switching between sleeping and active cells is observed
in our simulation when the balance between the numbers of active and sleeping
cells is destroyed. There are two origins for this destruction;
(a) internal dynamics (typically seen at the ``complex-III" stage
where the dynamics of many active cells is chaotic);
(b) external perturbation:
The second case is seen, when, for example, some cells are eliminated
externally. Then some sleeping cells become active.
Then some sleeping cell becomes active, and the
balance is restored, implying that our system has stability against external
perturbation.  In Fig. 9, the active cell is removed around $t=2030$.
After this removal, sleeping cells get active and compete for
resources.  After competition over few periods of oscillations,
one of the sleep cells wins and remains active.  Thus the original
state is restored. This stability might remind one of
``homeostasis".  However, the stability here is sustained not
by a static (fixed-point) state, but by a dynamical state.
Such dynamic stability with the use of high-dimensional chaos
is noted as homeochaos \cite{TIKK91a}.
\vspace{.1in}
----------------Fig. 9  ------------------
\vspace{.1in}
\item{ Stability of cell linkage}
Since chaotic dynamics underlies our cell system, one might
be afraid that the scenario here is unstable, sensitive to
initial conditions.  This is not the case.
The scenario here is, for example, invariant under
changes of initial conditions, ( except for a variation of
the time required for the first division), and also under
slight changes of parameters.  The oscillation itself
can be chaotic ( in stage II), and the time
for the division is not exactly identical, by the initial
conditions.  Still the division time
is statistically invariant, and the cell linkage is completely identical.
Of course, the scenario depends on genetic and environmental parameters,
such as the external supply of resources and internal parameters characterizing
chemical reactions in the cell.  Since the supply of chemicals from the
medium cannot be genetically determined, the genetic information is
not enough for the characterization of cell differentiation.
We should again emphasize here that spatial information is not necessarily
important for differentiation, either.
Examples of cell linkage diagrams are shown in Fig.10.
In Fig.10a), coherent division (stage I) is observed up to
32 cells ( at around 500 steps), while the cell society enters into
the stage III at around 600 steps. Here cells with an index larger than 107
get inactive (sleep cells).  Cells with the index less than 108 are
active and divide frequently but one of the created cells dies.
Around the time step 2200, divisions of many sleeping cells
occur leading to the multiple deaths.  Then the system goes back to
stage II, till new grouping into active and sleeping cells is organized
around 2700 steps.
On the other hand, the linkage of Fig.10b), corresponding to Fig.7,
shows the switching between active and sleeping cells around time steps 2000
and 4600.
\vspace{.2in}
------------Fig.10-------------------
\item{ Cell Death}
In our simulations with the deterministic cell death condition,
some cells are programmed to die.  This history of cell death is
stable against changes of initial conditions.
We have often observed simultaneous deaths of multiple cells.
The process eliminates the overgrowth of inactive cells in our simulation.
Such programmed death is also known in cell biology and immunology.
Indeed some of the cell linkages obtained (see Fig. 10) from our simulations
agree with those found for some multicellular organisms,
(such as the C elegance) in the following points.
(i) Not all of the cells divide (emergence of sleeping cells).  The number
of such cells and the cell linkage diagram are insensitive to
a wide-range change of initial condition.
Many cells are derived from few active cells.
(ii) The timing of divisions and
deaths of cells is independent of initial conditions.
If our scenario is true for the programmed death, we have to conclude
that programmed death is due to the interaction among cells.
It is predicted that the cell linkage by a single cell ( with removal
of one of divided cells) {\sl does not} show the programmed death
as seen in a cell society.  Experimental check for this conjecture
is strongly requested.
\end{itemize}
\section{Open Chaos; novel mechanism of unstable and irregular dynamics}
Although the mechanism in our differentiation is based on
dynamic clustering in a network of chaotic elements, there is
a novel dynamic instability in our system with ``growing" phase space.
By the active transport dynamics of chemicals, the
difference between two cells can be amplified, since a
cell with more chemicals is assumed to get even more.
Tiny differences between cells can grow exponentially
if parameters satisfy a suitable condition.  The grown cell
is divided into two, with an (almost) equal partition of the contained
chemicals.  The process is quite analogous with chaos in
Baker's transformations; stretching (exponential growth) and
folding (division).
One difference between our cell division mechanism
and chaos is that phase space itself changes after a division in
the former, while the orbit comes back to the original phase space
in the stretching-folding mechanism of chaos.
Thus conventional quantifiers ( such as Lyapunov exponents)
in "closed" chaos may not be applicable to
our problem, since they require a stationary measure in the
closed phase space.  A quantifier to measure the instability
along the increase of phase space is required just as
the co-moving Lyapunov exponent was introduced to measure the
instability along a flow \cite{D-KK}.
Our ``open chaos" here provides a general scenario of instability and
irregular dynamics in a system with growing phase space.  Such
systems with replicating units for the competition of finite resources
are often seen in economics and sociology.
In open chaos, disparity of elements in activities emerges through
time sharing by clustering.  Still, ``poor" cells are not extinct but
coexist.  Such coexistence of very active and inactive elements is also
seen in economics and sociology; coexistence of
very big and small firms, or the very rich and the poor.
Spatial information is not essential to the differentiation or class
formation here, as in our cell differentiation.
It is interesting to
extend the idea of the present paper to economics and sociology, and
to discuss the origin of differentiation, diversity, and complexity there.
\vspace{.1in}
The authors would like to thank F. Willeboordse for critical reading of the
manuscript.  This work is partially supported by a Grant-in-Aid for Scientific
Research from the Ministry of Education, Science, and Culture
of Japan.

\pagebreak
Figure Captions;
Fig.1  Schematic representation of our model: a) the whole
dynamics of our system; b)
metabolic reaction within each cell.
\vspace{.1in}
Fig.2  Temporal change of the number of cells $N$.   Throughout the present
paper we use $p=1$, $f=0.1$, $E^0=e_1=1$, and $R=50$.
Other parameters are set at
$ D=0.1$, $\delta =0.5$, $\overline{X^0}=15$, $S=0.1$.
All figures of the present paper  are plotted per $\Delta t =0.5$.
\vspace{.1in}
Fig.3   Overlaid time series of $x^{(2)} _i (t)$, corresponding to Fig.2.
\vspace{.1in}
Fig.4   (a) Plots of $(x^{(1)} _i (t),x^{(2)} _i (t))$ over the time steps
800-1000 ( stage II) (b)  The plots over the time steps 1900-2200
(stage III).  Corresponding to Fig.3.
\vspace{.1in}
Fig.5   Magnified plot of the time series $x^{(2)} _i (t)$,
for stage III ( corresponding to Figs. 2-4).
\vspace{.1in}
Fig.6 a) Temporal change of the number of cells $N$, for
the parameters $ D=0.05$, $\delta =0.5$, $\overline{X^0}=15$, $S=0.1$.
( for reference the evolution with $\overline{X^0}=35$ is plotted
as a dotted line).
b) Overlaid time series of $x^{(2)} _i (t)$, corresponding to Fig. 6a).
c) Plots of $(x^{(1)} _i (t),x^{(2)} _i (t))$ over the time steps 4000-4600.
\vspace{.1in}
Fig.7 Overlaid time series of $x^{(2)} _i (t)$, for the parameters
$ D=0.2$, $\delta =0.5$, $\overline{X^0}=15$, and $S=0.5$. At the time step
around 4620, switching between active and sleeping cells is observed.
\vspace{.1in}
Fig.8 Cell linkage diagram for $ D=0.1$, $\delta =0.5$, $\overline{X^0}=40$,
and $S=0.3$.  The vertical axis shows the time step,
while the horizontal axis shows the cell index.  ( Only for
practical purpose of keeping trade of the branching tree, we define
the cell index as follows: when a daughter cell $j$ is born from a cell $i$'s
$k$-th division, the value $s_j =s_i + 2^{-k}$ is attached to the cell $j$
from the mother cell's $s_i$.
The cell index for the cell $j$ is the order of $s_j$,
sorted with the increasing order).
In the diagram, the horizontal line shows the division from the
cell with index $n_i$ to $n_j$, while the vertical line
is drawn as long as the cell exists (until it dies out).
\vspace{.1in}
Fig.9 Overlaid time series of $x^{(2)} _i (t)$, for the parameters
$ D=0.1$, $\delta =0.5$, and $\overline{X^0}=40$.  Instead
of the deterministic death process, the stochastic decimation of
a cell is adopted.  A cell is randomly
removed with a probability of $10^{-5}$ per 0.01 second.
Around step 2050, the active cell is removed.
\vspace{.1in}
Figh.10 Cell linkage diagram (a)  $ D=0.05$, $\delta =0.3$,
$\overline{X^0}=15$, and $S=0.5$, up to time step 2910.
(b) $ D=0.2$, $\delta =0.5$, $\overline{X^0}=15$, and $S=0.5$, corresponding
to Fig.7; up to time step 4910.
\end{document}